\documentclass[prd,superscriptaddress,nofootinbib]{revtex4}
\usepackage{bm,amsmath,amssymb,epsfig,natbib}
\usepackage{hyperref}
\usepackage{ifthen}

\begin{document}
\newcommand{\nrun}{n_{\text{run}}}
\newcommand{\Neff}{N^{\text{eff}}}
\newcommand{\nueff}{\nu_{\text{eff}}}
\newcommand{\nside}{N_{\text{side}}}
\newcommand{\arcmin}{\text{arcmin}}
\newcommand{\begm}{\begin{pmatrix}}
\newcommand{\enm}{\end{pmatrix}}
\newcommand{\threej}[6]{{\begm #1 & #2 & #3 \\ #4 & #5 & #6 \enm}}
\newcommand{\Coff}{{\hat{C}_l^\off}}
\newcommand{\lmax}{l_\text{max}}
\newcommand{\lmin}{l_{\text{min}}}
\newcommand{\fsky}{f_{\text{sky}}}
\newcommand{\off}{{\text{off}}}
\newcommand{\chieff}{\chi^2_{\text{eff}}}
\renewcommand{\ell}{l}
\renewcommand{\bar}[1]{#1}
\newcommand{\nHI}{{n_{HI}}}
\newcommand{\clh}{\mathcal{H}}
\newcommand{\ud}{{\rm d}}
\newcommand{\fskyeff}{f_{\text{sky}}^{\text{eff}}}
\def\eprinttmp@#1arXiv:#2 [#3]#4@{
\ifthenelse{\equal{#3}{x}}{\href{http://arxiv.org/abs/#1}{#1}}{\href{http://arxiv.org/abs/#2}{arXiv:#2} [#3]}}

\renewcommand{\eprint}[1]{\eprinttmp@#1arXiv: [x]@}
\newcommand{\adsurl}[1]{\href{#1}{ADS}}
\renewcommand{\bibinfo}[2]{\ifthenelse{\equal{#1}{isbn}}{%
\href{http://cosmologist.info/ISBN/#2}{#2}}{#2}}

\newcommand\ba{\begin{eqnarray}}
\newcommand\ea{\end{eqnarray}}
\newcommand\be{\begin{equation}}
\newcommand\ee{\end{equation}}
\newcommand\lagrange{{\cal L}}
\newcommand\cll{{\cal L}}
\newcommand\clx{{\cal X}}
\newcommand\clz{{\cal Z}}
\newcommand\clv{{\cal V}}
\newcommand\clo{{\cal O}}
\newcommand\cla{{\cal A}}
\newcommand{\uD}{{\mathrm{D}}}
\newcommand{\calE}{{\cal E}}
\newcommand{\calB}{{\cal B}}
\newcommand{\curl}{\,\mbox{curl}\,}
\newcommand\del{\nabla}
\newcommand\Tr{{\rm Tr}}
\newcommand\half{{\frac{1}{2}}}
\renewcommand\H{{\cal H}}
\newcommand\K{{\rm K}}
\newcommand\mK{{\rm mK}}
\newcommand{\clk}{{\cal K}}
\newcommand{\bq}{\bar{q}}
\newcommand{\bv}{\bar{v}}
\renewcommand\P{{\cal P}}
\newcommand{\numfrac}[2]{{\textstyle \frac{#1}{#2}}}
\newcommand{\la}{\langle}
\newcommand{\ra}{\rangle}
\newcommand{\rar}{\rightarrow}
\newcommand{\Rar}{\Rightarrow}
\newcommand\gsim{ \lower .75ex \hbox{$\sim$} \llap{\raise .27ex \hbox{$>$}} }
\newcommand\lsim{ \lower .75ex \hbox{$\sim$} \llap{\raise .27ex \hbox{$<$}} }
\newcommand\bigdot[1] {\stackrel{\mbox{{\huge .}}}{#1}}
\newcommand\bigddot[1] {\stackrel{\mbox{{\huge ..}}}{#1}}
\newcommand{\Mpc}{\text{Mpc}}
\newcommand{\Al}{{A_l}}
\newcommand{\Bl}{{B_l}}
\newcommand{\eAl}{e^\Al}
\newcommand{\ix}{{(i)}}
\newcommand{\ixp}{{(i+1)}}
\renewcommand{\k}{\beta}
\newcommand{\HD}{\mathrm{D}}
\newcommand{\mCh}{\hat{\bm{C}}}

\newcommand{\mCf}{{{\bm{C}}_{f}}}
\newcommand{\mCXf}{{{\bm{C}}_{Xf}}}
\newcommand{\mMXf}{{{\bm{M}}_{f}}}
\newcommand{\cov}{\text{cov}}
\newcommand{\Cfl}{{C_f}_l}
\renewcommand{\vec}{\text{vec}}
\newcommand{\muK}{\mu \rm{K}}

\newcommand{\vecp}{\text{vecp}}
\newcommand{\vecL}{\text{vecl}}

\newcommand{\mCfl}{{\mC_{f}}_l}
\newcommand{\mCgl}{{\mC_{g}}_l}

\newcommand{\Ch}{\hat{C}}
\newcommand{\Bt}{\tilde{B}}
\newcommand{\Et}{\tilde{E}}
\newcommand{\bld}[1]{\mathrm{#1}}
\newcommand{\mLambda}{\bm{\Lambda}}
\newcommand{\mA}{\bm{A}}
\newcommand{\mB}{\bm{B}}
\newcommand{\mBp}{\mB_n}

\newcommand{\mC}{\bm{C}}
\newcommand{\mD}{\bm{D}}
\newcommand{\mE}{\bm{E}}
\newcommand{\mF}{\bm{F}}
\newcommand{\mg}{\bm{g}}

\newcommand{\mQ}{\bm{Q}}
\newcommand{\mU}{\bm{U}}
\newcommand{\mX}{\bm{X}}
\newcommand{\mV}{\bm{V}}
\newcommand{\mP}{\bm{P}}
\newcommand{\mR}{\bm{R}}
\newcommand{\mW}{\bm{W}}
\newcommand{\mI}{\bm{I}}
\newcommand{\mH}{\bm{H}}
\newcommand{\mM}{\bm{M}}
\newcommand{\mN}{\bm{N}}
\newcommand{\mMh}{\hat{\mM}}

\newcommand{\mS}{\bm{S}}
\newcommand{\mzero}{\bm{0}}
\newcommand{\mL}{\bm{L}}
\newcommand{\btheta}{\bm{\theta}}
\newcommand{\bphi}{\bm{\psi}}
\newcommand{\va}{\mathbf{a}}
\newcommand{\vX}{\mathbf{X}}
\newcommand{\vchi}{\bm{\chi}}
\newcommand{\vC}{{\mathbf{C}}}
\newcommand{\vv}{{\mathbf{v}}}
\newcommand{\vw}{{\mathbf{w}}}
\newcommand{\vk}{{\mathbf{k}}}
\newcommand{\vx}{{\mathbf{x}}}
\newcommand{\vn}{{\mathbf{n}}}
\renewcommand{\vr}{\mathbf{r}}

\newcommand{\vCh}{\hat{\mathbf{C}}}
\newcommand{\vCt}{\tilde{\mathbf{C}}}
\newcommand{\vNt}{\tilde{\mathbf{N}}}

\newcommand{\vXh}{\hat{\vX}}

\newcommand{\vS}{\mathbf{S}}
\newcommand{\vm}{\mathbf{m}}

\newcommand{\vN}{\mathbf{N}}
\newcommand{\vXhat}{\hat{\mathbf{X}}}
\newcommand{\vb}{\mathbf{b}}
\newcommand{\vA}{\mathbf{A}}
\newcommand{\vAt}{\tilde{\mathbf{A}}}
\newcommand{\ve}{\mathbf{e}}
\newcommand{\vE}{\mathbf{E}}
\newcommand{\vB}{\mathbf{B}}
\newcommand{\vl}{\mathbf{l}}
\newcommand{\vp}{\mathbf{p}}
\newcommand{\vXf}{\mathbf{X}_f}
\newcommand{\vEt}{\tilde{\mathbf{E}}}
\newcommand{\vBt}{\tilde{\mathbf{B}}}
\newcommand{\vEw}{\mathbf{E}_W}
\newcommand{\vBw}{\mathbf{B}_W}
\newcommand{\vXt}{\tilde{\vX}}
\newcommand{\vXb}{\bar{\vX}}
\newcommand{\vTb}{\bar{\vT}}
\newcommand{\vTt}{\tilde{\vT}}
\newcommand{\vY}{\mathbf{Y}}
\newcommand{\vBwr}{{\vBw^{(R)}}}
\newcommand{\RW}{{W^{(R)}}}
\newcommand{\mUt}{\tilde{\mU}}
\newcommand{\mVt}{\tilde{\mV}}
\newcommand{\mDt}{\tilde{\mD}}

\newcommand{\healpix}{HEALPix}
\newcommand{\Ctot}{C^{\text{tot}}}
\newcommand{\Chtot}{\hat{C}^{\text{tot}}}

\title{Cosmological parameters from WMAP 5-year temperature maps}

\author{Antony Lewis}
\homepage{http://cosmologist.info}
\affiliation{Institute of Astronomy, Madingley Road, Cambridge, CB3 0HA, UK.}

\date{\today}

\begin{abstract}
I calculate a hybrid cross-power spectrum estimator from the WMAP 5-year CMB temperature maps, discuss the goodness of fit, and then constrain cosmological parameters. The spectrum and results are generally consistent with previous results, though the power spectrum error bars are slightly smaller and there are small shifts at high $l$. The small improvement in error bars is obtained at very low numerical cost but does not significantly improve parameter constraints. I discuss the accuracy of the likelihood model and how constraints on the optical depth translate into constraints on the reionization history allowing for helium reionization. In the appendices I propose a simple reionization parameterization that determines the history in terms of a mid-point reionization redshift, and suggest a new likelihood approximation for chi-squared-like distributions with varying skewness.
\end{abstract}

\maketitle

\pagenumbering{arabic}

\section{Introduction}

With five years of data the WMAP satellite can measure the microwave sky down to sub-degree scales~\cite{Hinshaw:2008kr}, providing accurate measurements of the angular power spectrum and large scale polarization signal, and hence constraining many cosmological parameters~\cite{Dunkley:2008ie,Nolta:2008ih,Komatsu:2008hk}.
In this paper I analyse the temperature maps provided by the WMAP team, compressing the information into a set of power spectrum estimators, and then investigate the constraints on cosmological parameters. I use a hybrid cross-map power spectrum estimator that has slightly lower noise than the one used by the WMAP team. The effect on the cosmological parameters is modest,
and the final parameter values are very consistent with the WMAP 5-year results; I'm submitting this paper to help avoid any publication bias towards results that happen to drift in a particular direction.

Note that the scope of this paper is limited to a re-analysis of the foreground-cleaned temperature maps. I do not attempt to study the polarization signal or foreground removal. Details of the large scale polarization signal also impact parameters via the constraint on the optical depth, but otherwise all the interesting information about standard model parameters is in the temperature spectrum which is the easy part to analyse and the focus of this paper. I shall assume standard isotropic Gaussian cosmological models, that the WMAP likelihood code provides accurate results for the power spectrum at $l\alt 30$, the foreground model is correct, pixel noise is Gaussian and independent between maps and pixels, and that the beam model is accurate. This last assumption is perhaps the greatest for analysis of the temperature spectrum, with results depending critically on knowing the beam transfer function to sub-percent level. However the WMAP team have now studied the beams in great detail, and claim remaining uncertainties well below one percent~\cite{Hill:2008hx}. Uncertainties due to pixelization are hard to quantify other than by doing many full timestream simulations, and potentially important if the effect on the window function correction for pixel averaging is above a few percent. However consistency between ecliptic and galactic maps suggests that as expected these effects are small. The point source contribution to the spectrum is determined at the $10\%$-level in the five-year results, which is small enough for uncertainties to have only a minor effect on parameter constraints~\cite{Nolta:2008ih}; I assume the same model.

\section{Power spectrum estimation}

The general philosophy of most methods of small scale CMB data analysis is to first compress the data into a set of power spectrum estimators, then calculate the parameter likelihood given these estimators~\cite{Tegmark:1996qt,Wandelt:2000av,Hivon:2001jp,Hinshaw:2003ex}. Doing this potentially makes the analysis very fast, especially compared to a numerically very expensive direct likelihood evaluation from the timestream or maps. Depending on the choice of estimator, the amount of information lost by doing the compression can vary considerably. There is an extensive literature on different estimators, and like the WMAP team I shall concentrate on pseudo-$C_l$ estimators~\cite{Tegmark:1996qt,Wandelt:2000av}. These are especially fast to compute, allowing for extensive testing and Monte Carlo simulation for assessing errors.

The simplest Pseudo-$C_l$ estimator is constructed by taking some weight function $W(\Omega)$ on the sky, and then calculating the power spectrum of $W(\Omega)T(\Omega)$, where $T(\Omega)$ is the CMB temperature at position $\Omega$. The window function can be chosen to be zero over regions where the signal is not reliable (such as the galaxy and resolved point sources), but can otherwise be chosen in various ways. The expectation of the pseudo power spectrum $\tilde{\vC}$ is related to the true power spectrum $\vC$ by some coupling matrix $\mM$ and a noise power spectrum $\tilde{N}_l$,
\begin{equation}
\la \tilde{C}_l \ra = \sum_{l'} M_{l l'} C_{l'} + \tilde{N}_l.
\end{equation}
Hence by inverting $\mM$ one can construct unbiased pseudo-$C_l$ power spectrum estimators $\vCh = \mM^{-1} (\vCt - \vNt)$.
In practice there is more than one map, and different maps may have different noise and beam smoothing. Analysing a set of maps therefore requires some generalization. Methods for combining pseudo-$C_l$ estimators from different maps are well known and straightforward~\cite{Hinshaw:2003ex,Efstathiou:2003dj}: simply use the covariance between the estimators to work out a single combined minimum-variance estimator. This was done with the first-year WMAP results, using three different window functions $W(\Omega)$ for large, intermediate, and small scales; however different weight functions were not combined at any given $l$.  In the 5-year analysis the window function was taken to be uniform at $l < 500$, and inverse-noise on smaller scales; this choice is optimal at both very large and very small scales, but suboptimal on intermediate scales where the size of the error bars matters most. By combining weight functions smaller error bars can be obtained with essentially no extra computational effort~\cite{Efstathiou:2003dj}, in addition to giving aesthetically more pleasing error bars that are continuous in $l$.

In general one can define a set of weight functions $W^{(i)}(\Omega)$, and apply them to the set of available (foreground-cleaned) maps $T_\alpha(\Omega)$ to give the set of pseudo power spectra
\begin{equation}
\tilde{C}_{l,{\alpha\beta}}^{(ij)} \equiv \frac{1}{2l+1} \sum_m [\tilde{a}_{lm,\alpha}^{(i)} ][\tilde{a}_{lm,\beta}^{(j)}]^*,
\end{equation}
where
\begin{equation}
\tilde{a}_{lm,\alpha}^{(i)}\equiv \int \ud\Omega\, T_\alpha(\Omega) W^{(i)}(\Omega) Y_{lm}^*(\Omega),
\end{equation}
and $Y_{lm}$ is a spherical harmonic. The covariance of these estimators can be calculated approximately analytically, as given explicitly in the appendix of~\cite{Hamimeche:2008ai} for single maps.

When there there are multiple maps it is often an excellent approximation to assume their noise is uncorrelated. In the case of the WMAP 5-year data there are 4 detectors\footnote{I use `detector' as a shorthand for `differencing assembly'.} in W-band and 2 in V-band, giving a total of $(4+2)\times 5 = 30$ independent maps (other channels are not used as the beams are larger and there is significant foreground contamination). Each detector is assumed to have the same effective beam for each year of observation, though the noise varies slightly between years (for example due to different data cuts for planets, etc.). Given a set of estimators $\{\tilde{C}_{l,{bc}}^{(ij)}\}$ an optimal analysis would use all of them, using all prior information about the noise to model noise biases. However if there is significant uncertainty in the noise in each map, marginalizing out this uncertainty essentially down weights all spectra involving only one map. The WMAP team therefore sensibly just throw out all estimators with $\alpha=\beta$, giving a set of cross-map estimators that have zero noise bias:
\begin{equation}
(1-\delta_{\alpha\beta}) \la \vCh^{(ij)}_{\alpha\beta} \ra= (1-\delta_{\alpha\beta}) \la b_\alpha^{-1}b_\beta^{-1} [\mM^{ij}]^{-1} \vCt^{(ij)}_{\alpha\beta}\ra = (1-\delta_{\alpha\beta})\vC.
\end{equation}
The coupling matrix $\mM^{(ij)}$ now depends on the two weight functions, and the transfer functions $b_l$ take out the smoothing on each map due to the beam and pixelization (I assume the effective beam and pixel transfer function is the same over the masked map as over the full sky). If there are $n$ maps, the maximum increase in error bar from using only cross-spectra is $\sim 1/2n$, so for $30$ maps the loss is at the percent level: a small price to pay for cleanly removing most systematic errors relating to noise bias.

It remains to choose the weight functions $\{W^{(i)}(\Omega)\}$. On large scales the noise is very small, and errors are limited by cosmic variance, and by symmetry the optimal weighting must be uniform on the full sky. On small scales noise dominates, and the weight function should give more weight to parts of the sky observed more often (and hence having lower noise). Inverse-noise weighting is the optimal choice for a single map~\cite{Efstathiou:2003dj}. I also use the `KQ85' mask to remove foreground contamination (setting to zero $\sim 15\%$ of the weight map); this is smoothed on a $1/3$-degree scale to reduce $l$-couplings without significantly degrading the result.\footnote{Since the CMB power spectrum falls rapidly at $l\agt 200$, mode coupling tends to increase the variance at a given $l$ by mixing in large-scale modes with larger variance. Apodizing improves the accuracy of the approximations used to estimate the $C_l$ covariance, with further improvements to the few-percent level obtained by first smoothing the fiducial power spectrum on a scale appropriate to the weight function.} I therefore choose to use two weight functions, ``$KQ85 \times \text{uniform}$'' and ``$KQ85\times\text{inverse noise}$'', where the `inverse-noise' map is calculated from a smoothed combined map. An optimal combination of estimators using these weights should smoothly interpolate between small and large scales where the weight functions are individually close to optimal. For further discussion of how to optimally choose a weight function for a $C_l$-estimator at a particular $l$ from one map see Ref.~\cite{Smith:2006hv}.

\begin{figure}
\begin{center}
\epsfig{figure=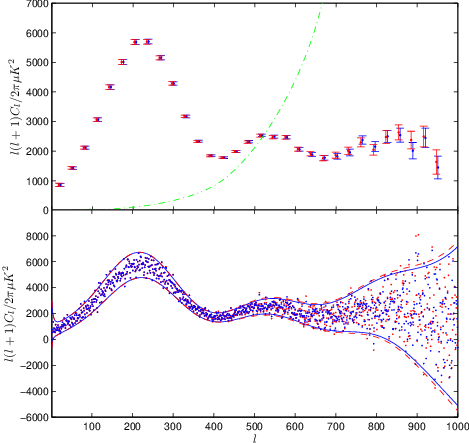,angle=0,width=12cm}
\caption{The hybrid power spectrum (blue) compared with the WMAP estimators (red). The top panel is binned with $\Delta_l=31$, and error bars show $\pm 1\sigma$ error bars slightly offset for clarity; the dash dotted line is $\Neff_l$. The lower panel shows the estimators at each $l$, with the lines showing $2\sigma$ symmetric posterior diagonal errors from the assumed fiducial model; the slightly broader dashed lines are the corresponding diagonal errors from the WMAP likelihood code.
}
\label{spectrum}
\end{center}
\end{figure}

Putting this all together the hybrid cross-estimator is just a particular linear combination of all the separate estimators:
\begin{equation}
\vCh = \sum_{ij\alpha\beta} H^{(ij)}_{\alpha\beta} (1-\delta_{\alpha\beta})\vCh^{(ij)}_{\alpha\beta}.
\end{equation}

With $n_{\hat{C}_l}$ distinct estimators and $n_l$ different $l$-values, calculation of $H^{(ij)}_{\alpha\beta}$ requires inversion of the full $[n_l\times n_{\hat{C}_l}]^2$ covariance matrix of the different estimators. This becomes computationally tedious or even difficult if there are many maps. Clearly there is no point in using an approximate method that is becoming as expensive as doing a proper lossless likelihood analysis. One option would be to neglect some of the very small off-diagonal correlations. Here, instead of calculating the full result for the full set of maps, I find the optimal mixing for combined-channel maps. Since the noise is very similar between years, and the beams for each detector in each band are quite similar, the dominant variation of the hybrid matrix comes from the weight and band index, rather than variations between years or detectors. Since there are many maps the difference between cross and auto-power spectrum estimator covariances is small.  I then only need to calculate the covariance of $\vCh^{(ij)}_{AB}$, where $A$ and $B$ label some combination of the $V$ and $W$ bands (giving a total of 10 distinct power spectrum estimators). The combined maps can be calculated easily by co-adding the maps from different years\footnote{The word `co-add' appears to be specific to the cosmological data analysis community. I use a `co-added' set of vectors $\{\vv^{(i)}\}$ to mean another vector $\vw$ of the form $w_j = \sum_i \alpha^{(i)}_j v^{(i)}_j$, where the weights $\alpha^{(i)}_j$ are chosen to approximately minimize the noise on $w_j$.}, and then summing the maps from the different detectors. This last step is slightly suboptimal due to the differences in beams and noise between detectors; in fact for WMAP this suboptimality is almost the same as that from using cross-spectrum estimators, with the result that the hybrid estimator covariance calculated from combined-map Monte Carlo agrees at the sub-percent level (on the diagonals) with that from the hybrid estimator applied to co-added cross-spectra (see Appendix~\ref{app:like}).

Rather than taxing the index-tolerance of the reader any further by explicitly writing down how the full combined-map covariance (and hence $ H^{(ij)}_{\alpha\beta}$) is computed, I simply refer to the appendix of Ref.~\cite{Hamimeche:2008ai} and Refs.~\cite{Hinshaw:2003ex,Efstathiou:2003dj} from which the result is a straightforward generalization.  Once the hybrid mixing matrix is computed, one full map-set simulation and cross-power spectrum estimation with two weight maps takes about three minutes over 8 processors at `res 10' ($12\times 1024^2$ \healpix\ pixels) and using $\lmax =1100$. Simulation and analysis of combined-map hybrid estimators is significantly faster since only four weighted maps have to be spherical-transformed (two each in the $V$ and $W$ bands) rather than 60. The full cross-spectrum analysis of the WMAP maps involves estimating the noise, calculating combined maps, generating the mixing matrix and covariance for all the estimators, forming the combined-map hybrid mixing matrix, calculating the hybrid estimator covariance, and then calculating the hybrid cross-power spectrum from the map data; all this takes around 20 minutes on 8 processors, making it fast enough to easily test the effect of changing various aspects of the analysis. The numerical code is available for public inspection and use\footnote{\url{http://cosmologist.info/weightmixer/}}, and the spectrum from the foreground-cleaned WMAP 5-year maps\footnote{\url{http://lambda.gsfc.nasa.gov/}} is shown in Fig.~\ref{spectrum}

The power spectrum is very consistent with the WMAP analysis where it is cosmic variance limited. Once the noise becomes significant the hybrid estimator performs slightly better. The error bars in this analysis remain $\sim 10\%$ smaller to small scales, and it also looks as though the outliers at high $l$ are more consistent with the error bars in this analysis than the official WMAP 5-year spectrum (e.g. see the group of points around $l=900$). The WMAP points are consistently slightly higher in the  high-$l$ noise dominated region, though the reason for this is not entirely clear. Differences in the very noise-dominated region have little effect on standard parameter constraints because the error bars are large compared to the range of models that fit the first two acoustic peaks.

\section{Calculating the likelihood}

Having compressed the map data into a vector of estimators $\vCh$, the next step is to calculate the likelihood $\cll(\vC(\theta)|\vCh)$, where $\vC(\theta)$ is the theoretical power spectrum from a set of cosmological parameters $\theta$. This step is again computationally prohibitive to do exactly, so some approximation is required --- for a detailed recent discussion  of possibilities see Ref.~\cite{Hamimeche:2008ai}. The approximation used by the WMAP likelihood code is described in Ref.~\cite{Verde:2003ey} and the latest treatment of beam and point source uncertainties in Refs.~\cite{Hinshaw:2006ia,Nolta:2008ih}. For the low-$l$ spectrum ($l\le 32$) I use the WMAP likelihood code as supplied, simply treating the high-$l$ temperature likelihood as an independent dataset. To isolate changes due to the new temperature map analysis I include the high-$l$ temperate-polarization correlation spectrum using the original WMAP likelihood code.

For the high-$l$ temperature likelihood I use the new approximation suggested in Ref.~\cite{Hamimeche:2008ai}:
\begin{equation}
-2\log\cll(\vC|\vCh) \approx \sum_{ll'} g\left(\frac{\hat{C}_l+\Neff_l}{C_l+\Neff_l}\right) \left(C_{f,l}+\Neff_l \right) [M_f^{-1}]_{ll'} \left(C_{f,l'}+\Neff_{l'} \right)
g\left(\frac{\hat{C}_{l'}+\Neff_{l'}}{C_{l'}+\Neff_{l'}}\right),
\label{like_approx}
\end{equation}
where $\mM_f$ is the estimator covariance evaluated for a fiducial model $C_{f,l}$, $\Neff_l$ is an $l$-dependent parameter to be chosen, and $g(x)\equiv\text{sign}(x-1)\sqrt{2(x-\log(x)-1)}$ accounts for the skewness of the chi-squared-like distribution of the $\vCh$ estimators. For a consistent choice of $\Neff_l$ and the same covariance this gives almost exactly the results as the WMAP likelihood approximation for the statistical errors (though a slightly different $\chi^2_{\text{eff}}$). Using the new likelihood is therefore a cross-check rather than an a significant improvement on the WMAP analysis.
I take
\begin{equation}
\Neff_l \equiv \left(\sqrt{M_{ll}^{(S+N)}/M_{ll}^{(S)}} -1 \right) C_{f,l},
\end{equation}
where $M^{(S+N)}_{ll'}$ is the estimator covariance and  $M^{(S)}_{ll'}$ is the covariance when there is no noise, both evaluated at the fiducial model $C_{f,l}$. For this purpose the covariances are calculated using the analytic approximations; if $\Neff_l$ is only used in Eq.~\ref{like_approx} accuracy is not critical. This choice of $\Neff$ is consistent with the WMAP likelihood model where the diagonals of the covariance are taken to scale with the noise as $\propto (C_l + \Neff_l)^2$. For further discussion and tests of the accuracy of the likelihood model see Appendix~\ref{app:like}.

Assuming point sources behave like an isotropic gaussian random field with a white spectrum, their contribution can be added to $C_{f,l}$ and $C_{l}$ when calculating the likelihood (with their cosmic variance contribution included in $\mM_f$).
I use a fiducial unresolved point source power spectrum with amplitude $A_{ps}=0.011\muK^2$ (following Ref.~\cite{Nolta:2008ih}, with spectral index parameter $\alpha=0$). I calculate the point source contribution $C_{p,l}$ to the $C_l$ by combining the assumed white-noise spectra for each frequency appropriately over the hybrid matrix. The estimators shown in Fig.~\ref{spectrum} have had this contribution subtracted. For further discussion of the point source contribution see Refs.~\cite{Huffenberger:2007sc,Wright:2008ib}. My treatment of point source and beam uncertainties is described in the next section.

\begin{figure}
\begin{center}
\epsfig{figure=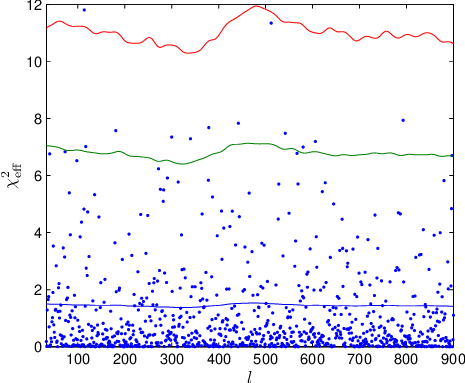,angle=0,width=10cm}
\caption{Diagonal contributions to the likelihood from each $l$ of the observed cross-spectrum $\Ch_l$. The lines show the smoothed rms, 0.01 and 0.001 confidence limits estimated from simulations of the combined map spectra (bottom to top). Points move around a bit depending on which model is used to calculate the contribution to the likelihood.
}
\label{chisqs}
\end{center}
\end{figure}
The effective chi-squared $\chieff$ (defined by Eq.~\eqref{like_approx}) of the power spectrum estimators is consistent with simulations, giving $\chieff=936$ (reduced value 1.08) using $33\le l\le 900$ for the fiducial model. This compares to $\chieff = 882\pm 44$ in simulations of the fiducial model, where all values are computed using the analytic approximation for the covariance and zero SZ contribution. The observed value is a little high, but this should not be surprising since the fiducial model is expected to be somewhat off the true model, and there are also the beam and point source uncertainties. In any case the value is consistent at about the 1-sigma level. Consistency of $\chieff$ values is only a very minimal consistency check: individual $\Ch_l$ could still vary significantly from the expected distribution. However Fig.~\ref{chisqs} shows that although there are apparent outliers in the spectrum, they only occur with a frequency about that expected by comparison with simulations. In the absence of any immediate evidence for deviations from the assumed theoretical framework on the relevant scales, I proceed to estimate parameters using the likelihood approximation described above.

\section{Parameter estimation}

\begin{figure}
\begin{center}
\epsfig{figure=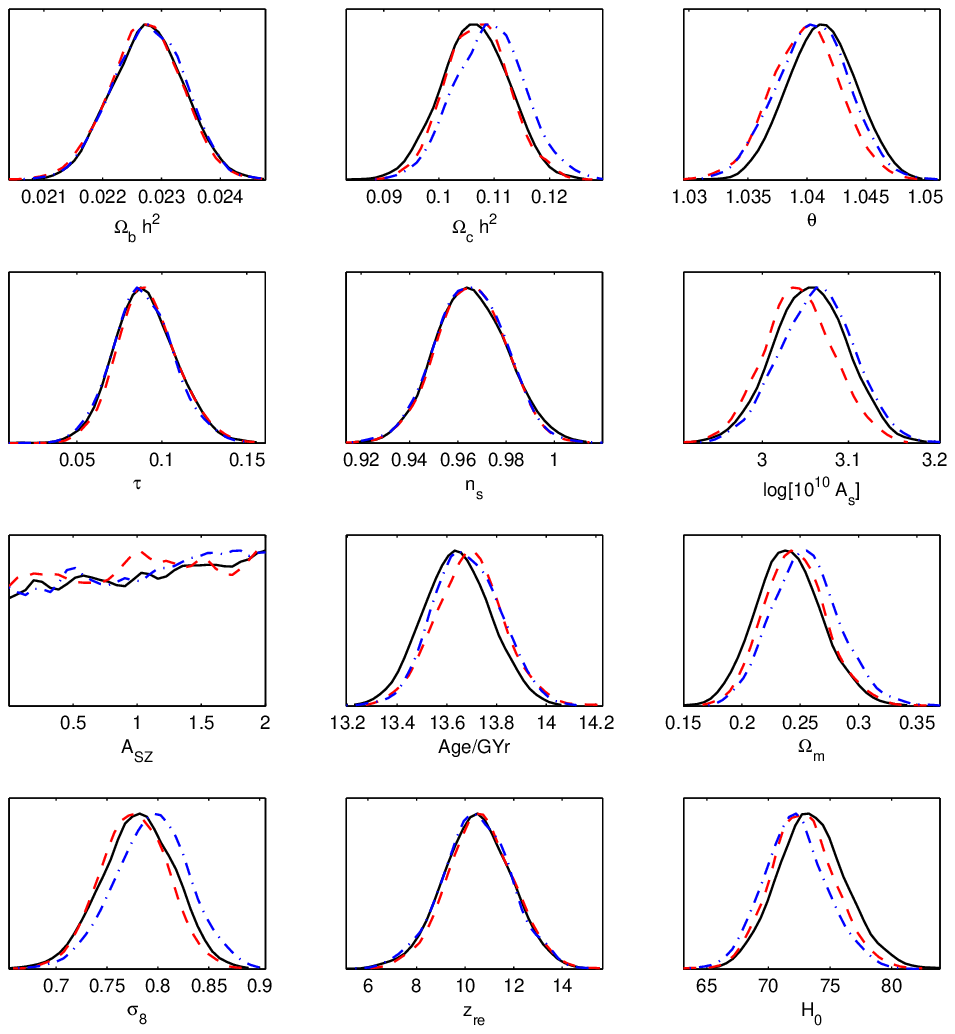,angle=0,width=10cm}
\caption{Marginalized parameter constraints of this paper (solid lines) compared to using the WMAP likelihood code with the spectrum and covariance from this paper (dashed) and the result from the original WMAP likelihood code and spectrum (dot-dashed). The bottom five parameters (except $A_{SZ}$) are derived from the other parameters.
}
\label{LAMBDA}
\end{center}
\end{figure}

Using the power spectrum estimators and model for the likelihood function, cosmological parameters can be sampled using standard Markov chain Monte Carlo methods. I use the CosmoMC\footnote{\url{http://cosmologist.info/cosmomc/}} code~\cite{Lewis:2002ah}, which uses CAMB~\cite{Lewis:1999bs} to calculate the theoretical power spectra, and restrict to vanilla flat $\Lambda$CDM models with six parameters of interest: reionization optical depth $\tau$, dark matter density $\Omega_c h^2$, baryon density $\Omega_b h^2$, amplitude of the primordial adiabatic curvature perturbation power spectrum $A_s$ at $k=0.05\Mpc^{-1}$ (with flat prior on $\log A_s$), constant spectral index $n_s$, and angular scale measured by $\theta$ --- approximately 100 times the ratio of the sound horizon to the angular diameter distance at last scattering. Other parameters such as the Hubble expansion rate $H_0$ today can be derived from these. I include the effect of CMB lensing and also marginalize over a SZ template amplitude $A_{SZ}$ as in the WMAP papers~\cite{Dunkley:2008ie,Komatsu:2008hk}. I generate six chains, stopping when the worst rms difference in 95\% parameter confidence limits between chains is a small fraction of the error bar.

In addition to the cosmological parameters and $A_{SZ}$ I include a point source spectrum amplitude parameter to account for the $10\%$ uncertainty in the point source power spectrum amplitude.  Accounting for beam uncertainties properly is more difficult without knowing the uncertainty for each detector. However the beams are now so accurately modelled that beam uncertainties are not a large effect; I therefore simply take the 10 beam uncertainty modes identified by the WMAP team and assume that they also apply to the hybrid power spectrum. I add mode amplitude parameters to the Markov chain with Gaussian priors, including the effect of the beam modes on the theoretical power spectrum at each point in parameter space. In practice only the most important modes have a significant effect, so I include the top five of the ten modes as Monte Carlo parameters, and add the remaining five modes as tiny non-diagonal terms in the covariance matrix. This gives essentially identical results to including all 10 beam modes in the chain. In total the Markov chain includes thirteen parameters: six cosmological parameters of most interest, $A_{SZ}$, one point source amplitude parameter and five beam modes. The nuisance parameters are prior driven, so their posteriors are similar to their priors.

To calculate the theoretical power spectrum from the set of cosmological parameters some shape of the reionization history has to be assumed. Many different histories can give the same optical depth. I parameterize the history by a redshift parameter $z_{re}$ and a fixed width $\Delta_z$,  see Appendix~\ref{sec:reion} for details. In this parameterization $z_{re}$ measures where the ionization fraction was half of its maximum. The maximum fraction is about $x_e\sim 1.08$ electrons per hydrogen atom assuming that hydrogen reionizes in roughly the same way as the first reionization of helium. The WMAP team used a different definition of $z_{re}$ --- relating it to a sharp reionization model with $x_e=1$ at $z\alt z_{re}$ --- so my values for this parameter differ by about 6\% from the WMAP 5-year results. Although this is primarily a difference of definition, results also differ slightly because of the slightly different large-scale $E$-polarization signal when the reionization history changes. The new reionization parameterization has been included in CAMB\footnote{\url{http://camb.info}} since the March 2008 version (with  $\Delta_z = 0.5$). For a more detailed investigation of details of the reionization history see Ref.~\cite{Mortonson:2008rx}.

Figure~\ref{LAMBDA} shows marginalized parameter constraints compared to those obtained from the WMAP likelihood code. The spectral index constraint is $n_s = 0.966\pm 0.015$ (quoting 1-sigma errors), and as before $n_s=1$ is ruled out at just over 2-sigma. The dark matter density has shifted slightly lower, with $\Omega_c h^2 = 0.106\pm  0.006$.
Other parameter values are also broadly consistent, with slight shifts to $\sigma_8 = 0.780 \pm 0.036$ and $\Omega_m = 0.24\pm 0.03$, and well constrained combination $\Omega_m h^{2.6} \sigma_8^{-0.62} = 0.125\pm 0.002$.
The redshift of reionization in the new parameterization is $z_{re} = 10.5\pm 1.4$, with a 2-sigma lower limit of $z_{re} > 7.8$. Extending the analysis to include a running of the spectral index $\nrun$ gives $\nrun = -0.037\pm 0.027$, very consistent with the WMAP result.

There is a slight shift in poorly determined parameters like $\Omega_m$ and the Hubble parameters apparent in Fig.~\ref{LAMBDA} when using the new spectrum, $\Neff$, covariance and point source spectrum in the WMAP likelihood code compared to the likelihood analysis described here. The reason for this is not entirely obvious, with a combination of small effects from different beam and point source uncertainty modelling, different likelihood approximation, and approximations used in the WMAP code. The shift may be a good indicator of the likely range of systematic error in the results due to likelihood modelling. These parameters are in any case somewhat sensitive to the choice of priors. The systematic error due to the low-$l$ likelihood can be estimated from Ref.~\cite{Dunkley:2008ie}, where changes in the polarization analysis shift the optical depth by $\sim 0.01$, comparable to the effect of reionization history modelling~\cite{Mortonson:2008rx}: I have not accounted for these uncertainties in my results.

\section{Conclusions}

I presented a re-analysis of the 5-year WMAP temperature maps. This was certainly not independent of the analysis done by the WMAP team, however it is very reassuring that the results are so consistent. Using an improved power spectrum estimator can reduce the information loss when compressing the sky maps into a set of estimators, but the ultimate parameter constraints are very similar. Future experiments such as the Planck satellite will have a much more anisotropic noise distribution, making the gain from using hybrid-like estimators much more significant. More optimal estimators may do even better and be worth calculating if the likelihood function for the estimators can still be calculated reliably.

The general methodology applied in this paper extends straightforwardly to small-scale polarization analysis as discussed and proven for single maps in Ref.~\cite{Hamimeche:2008ai}. I have not attempted a reanalysis of the WMAP T-E power spectrum here since it contains only a modest amount of information on smaller scales, though using a better estimator might shrink the error bars slightly. The new likelihood parameterization would also  allow the $l-l$ and $T$-$E$ correlations to be accounted for more consistently than in the current WMAP likelihood code. In the future a full consistent analysis of temperature and polarization will be much more worthwhile.

This analysis was possible because the WMAP team have made their data available. As a further step towards reproducibility my numerical code is publicly available, so my power spectrum results can be exactly reproduced from the data supplied by the WMAP team.

\section{Acknowledgements}
I acknowledge a PPARC/STFC Advanced fellowship and thank the WMAP team for making their data so easily accessible and well documented\footnote{\url{http://lambda.gsfc.nasa.gov/}}. Some of the results in this paper have been derived using the \healpix~\cite{Gorski:2004by} package. I thank the George Efstathiou, Steven Gratton, Hiranya Peiris and Kendrick Smith for discussion and comments, and members of the WMAP team for helpful communication.
\appendix

\section{Accuracy of likelihood model}
\label{app:like}

To model the likelihood accurately the approximation of Eq.~\eqref{like_approx} needs to be reliable, and the covariance (and effective noise) have to be sufficiently accurate.

Construction of the hybrid mixing matrix is done using analytic approximations for the covariances, which are accurate at the $\alt 20\%$-level, depending on the degree of apodization of the mask used. The main inaccuracy is in the signal part of the covariance, where it is effectively assumed that $C_l$ are constant over the range of $l$ that are coupled by the sky cut and weighting. The noise can be calculated accurately from the noise maps. This of course assumes the assumption of uncorrelated Gaussian noise is correct, and that noise variance is well known. I estimate the pixel-noise variance directly from the maps by assuming that the maps for different years differ only in their independent noise realizations. There is some evidence for drifts and variations in the noise variance at the $0.8\%$ level; this would be enough to be slightly worrying if including auto-spectra in the analysis, but when using only cross-spectra mis-estimation only has a very small effect on the error bars. Calculating cross-spectra from linear combinations of year maps that should be independent of the CMB does give a result consistent with pure noise simulations.

\begin{figure}
\begin{center}
\epsfig{figure=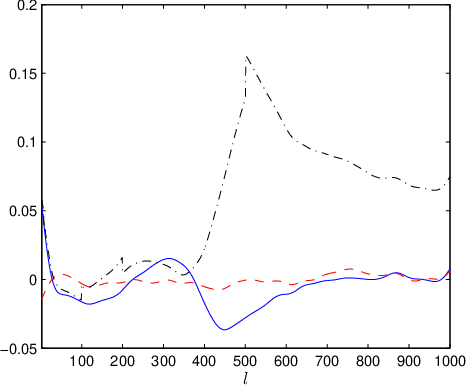,angle=0,width=8cm}
\caption{The fractional difference in diagonal error as a function of $l$ (i.e. the square root of the diagonal of the $\Ch_l$ covariance) compared to that from 1500 simulations of the hybrid cross-spectrum estimator.  The solid line shows the analytic approximation for the combined-map hybrid spectrum (having smoothed the input power spectrum with $\Delta_l\sim 25$ for $l>200$). The dashed line is from 5000 simulations of the combined-map hybrid estimator. The dash-dotted line shows the result used by the WMAP likelihood code supplied on LAMBDA (which is discontinuous at $l=500$ where they switch from uniform to inverse-noise weighting). Simulation results are smoothed over $\Delta_l=20$ to reduce sampling noise. The analytic result is accurate on small scales because the noise dominates.
}
\label{covdiffs}
\end{center}
\end{figure}

Figure~\ref{covdiffs} compares the diagonal values of the covariance for the various different cases and compares with the WMAP 5-year likelihood model. The diagonal cross-spectra and combined-map covariances agree very well, and agree with the analytic result to $< 10\%$. The WMAP error bars are consistently larger than from the hybrid estimators as expected when the signal and noise are both significant. It is not entirely clear why the results do not converge at high $l$, though inspection of the point-source power spectrum amplitudes suggests the WMAP mixing matrix is using a slightly different mix of $W$ and $V$ from the hybrid estimator.

In addition to having the covariance accurately, there is also a question about whether the likelihood model is reliable. If the non-Gaussianity of the estimator distributions is not correctly modelled it could potentially give parameter biases as well as giving wrong error bars. The likelihood approximation of Eq.~\eqref{like_approx} is essentially treating each $\Ch_l$-estimator as though it had a (reduced) chi-squared distribution with some degrees of freedom $\nu_l$.

As discussed in Ref.~\cite{Hamimeche:2008ai}, the correctness of the statistical error bars can be checked using binning: with wider bins the distributions become more Gaussian by the central limit theorem. Since most theoretical models give very smooth power spectrum (e.g. accurately recovered by splining points separated by $\Delta_l=50$ on small scales), the information lost by binning with $\Delta_l \ll 50$ is very small. I calculated chains using a fiducial Gaussian likelihood model and binned spectra with $\Delta_l=10$. In this limit beam and point source uncertainty modes can just be added to the covariance as terms of the form $\vC_m\vC_m^T$ for each mode $\vC_m$.
 Parameter constraints agree very well with those from un-binned estimators and including the beam and point source parameters in the Markov Chain. This provides some confirmation that the likelihood model is consistent. However the analysis does rely on the correct beam modes being identified; since I have only treated this approximately the WMAP analysis should be more reliable in this respect. A more sophisticated treatment could also account for the non-Gaussian and non-isotropic distribution of point sources.


\subsection{Skewness}

\begin{figure}
\begin{center}
\epsfig{figure=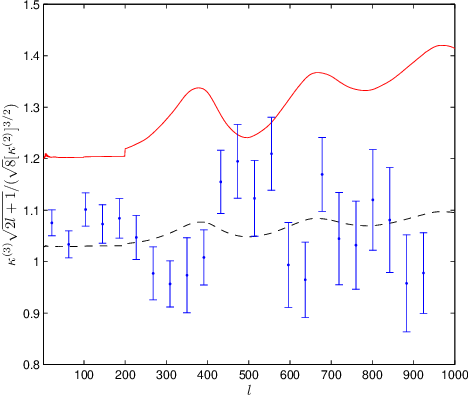,angle=0,width=8cm}
\caption{The skewness ($\kappa^{(3)}/[\kappa^{(2)}]^{3/2}$) as a fraction of that expected from a reduced chi-squared distribution on the full sky. The solid lines is the prediction for the hybrid estimator based on assuming a reduced chi-squared distribution determined by the variance and degrees of freedom. Points are binned estimates from 10000 simulations with error bars estimated from the values in each bin. The dashed line is a possible improved model with $\alpha = 1-\fskyeff$.
}
\label{skew}
\end{center}
\end{figure}

For the likelihood approximation to be good at each $l$ the estimators need to have a distribution that is a reduced chi-squared with some $\nu_l$ degrees of freedom. This is a very specific distribution where the skewness is fixed by the variance and effective degrees of freedom. Since the skewness is related to likelihood bias, any difference in skewness leads to errors in the likelihood approximation at each $l$. As shown in Ref.~\cite{Hamimeche:2008ai} accuracy at each $l$ is not actually required  for reliable parameter constraints: differences between $l$ tend to average out in many cases.
For example using a Gaussian approximation is actually reliable despite being a poor model at each $l$. Above we have confirmed that the likelihood approximation gives reliable parameter constraints for simple models, but in general it would be nice to have an approximation that is reliable $l$-by$-l$, for example when considering strange models with glitches in the spectrum.

Figure~\ref{skew} shows the skewness estimated from simulations. The skewness is significantly lower than expected from the estimator variance and effective degrees of freedom $\nu$ estimated by taking the diagonal covariance to be $2C_l^2/\nu$ when there is no noise. Expected small differences between auto- and cross-spectra~\cite{Hamimeche:2008ai} cannot be resolved without doing many more simulations. This motivates finding a likelihood approximation that can use a more accurate model of the skewness.

On the full sky with no noise the maximum likelihood estimators are $\Ch_l\equiv \sum_m |a_{lm}|^2/(2l+1)$, which have a reduced chi-squared distribution, so
\begin{equation}
-2P(\Ch_l|C_l) = \nu\left[ \Ch_l/C_l - \log (\Ch_l/C_l)\right] -\nu\log(\nu/2)+2\log\left[\Gamma(\nu/2)\right]
+ 2\log(\Ch_l),
\end{equation}
where $\nu=2l+1$ is the degrees of freedom. With isotropic noise the same result holds with $C_l\rightarrow C_l+N_l$ and $\Ch_l\rightarrow C_l+N_l$ since the noise is also assumed to be Gaussian. This motivates generally considering whether
\begin{multline}
-2P(\Ch_l|C_l) \approx \nueff\left[ \frac{\Chtot_l+\alpha\Ctot_l}{(1+\alpha)\Ctot_l} - \log\left(\frac{\Chtot_l+\alpha\Ctot_l}{(1+\alpha)\Ctot_l}\right) \right]
+ 2\log(\Chtot_l+\alpha \Ctot_l) \\
 -\nueff\log(\nueff/2)+2\log\left[\Gamma(\nueff/2)\right],
\label{skewdiag}
\end{multline}
where $\Ctot_l \equiv C_l+\Neff_l$, $\Chtot_l \equiv \Ch + \Neff_l$. This has $\la \Ch_l\ra = C_l$ as required for unbiased estimators, and $\nueff$ and $\alpha$ (generally both functions of $l$) could be obtained from the second and third moments. Defining $\nu\equiv \nueff/(1+\alpha)^2$ these are
\begin{eqnarray}
\kappa^{(2)}_l &\equiv& \la (\Ch_l-C_l)^2\ra = \frac{2(\Ctot_l)^2}{\nu},\\
\kappa^{(3)}_l &\equiv&   \la (\Ch_l-C_l)^3\ra = \frac{8(\Ctot_l)^3}{(1+\alpha)\nu^2}.
\end{eqnarray}
The parameter $\alpha$ controls the skewness relative to that expected for a chi-squared distribution with $\nu$ degrees of freedom. The limit $\alpha\rightarrow 0$ corresponds to the reduced chi-squared distribution, and $\alpha\rightarrow\infty$ gives
\begin{equation}
-2P(\Ch_l|C_l) \rightarrow \frac{\nu(C_l-\Ch_l)^2}{2[\Ctot]^2} + 2 \log(\Ctot_l)+\text{const},
\end{equation}
i.e. a Gaussian. Considered as a likelihood function the maximum is not at $C_l=\Ch_l$ unless $\alpha=0$: assuming $\nu \gg \alpha$ it is instead at $\approx \Ch_l[1-2\alpha/\nu(1+\alpha)+\dots]$. Using Eq.~\eqref{skewdiag} to fit for $\alpha$ from simulations gives values from about $0.15$ on large scales to $0.35$ on small scales, consistent with Fig.~\ref{skew}.

Generalizing Eq.~\eqref{skewdiag} to the cut sky suggests the likelihood approximation
\begin{multline}
-2\cll(\vC|\vCh) \approx   \sum_{ll'} g\left( \frac{\Chtot_l+\alpha_l\Ctot_l}{(1+\alpha_l)\Ctot_l}\right)(1+\alpha_l)\Ctot_l    [\mM^{-1}]_{ll'}\Ctot_{l'}(1+\alpha_{l'}) g\left( \frac{\Chtot_{l'}+\alpha_{l'}\Ctot_{l'}}{(1+\alpha_{l'})\Ctot_{l'}}\right) \\
+ 2\sum_l \log\left(\frac{\Chtot_l+\alpha_l \Ctot_l}{(1+\alpha_l)\Chtot_l}\right)
\end{multline}
where $g(x)\equiv\text{sign}(x-1)\sqrt{2(x-\log(x)-1)}$. Here terms independent of $C_l$ have been changed such that the normalization $\cll(\vCh|\vCh)=0$ (which is \emph{not} in general the maximum likelihood $\vC$). The variable $\Neff_l$ is defined so that $\Ctot_l    [\mM^{-1}]_{ll'}\Ctot_{l'}$ is almost independent of $\vC$.
This could be generalized for polarization following the vectorization method of Ref.~\cite{Hamimeche:2008ai}.

Using this likelihood approximation using the crude fitting $\alpha_l=1-\fskyeff{}_l$ (with $\fskyeff{}_l \equiv \sqrt{\nu/(2l+1)}$) gives virtually identical results for the parameters to Eq.~\eqref{like_approx}. This should be no surprise since using a Gaussian approximation also gives very similar results (as expected in typical realizations). If there were strongly $l$-dependent skewness parameters $\alpha_l$ not resolved with a small number of simulations the effect could be larger. The approximation may also be useful at low $l$ where $\Ch_l+\Neff_l$ has a significant probability to be negative using Pseudo-$C_l$ or quadratic maximum likelihood estimators.


\section{Reionization parameterization}
\label{sec:reion}
The optical depth to reionization is defined by
\begin{equation}
\tau = \int_0^{\eta_0} \ud\eta \,a n_e^{\text{reion}} \sigma_T,
\end{equation}
where $n_e^{\text{reion}}$ is the number density of free electrons produced by reionization  at conformal time $\eta$, $\eta_0$ is the conformal time today, $\sigma_T$ is the Thomson scattering cross-section, and $a$ is the scale factor. The total number density of free electrons is slightly different from $n_e^{\text{reion}}$ because there is a small residual ionization fraction from recombination. At the level of precision required this can be neglected ($\alt 10^{-3}$), though CAMB keeps the ionization history smooth my mapping smoothly onto the recombination-residual.

Since $n_e \propto (1+z)^3x_e(z)$, where $x_e$ is the number of free electrons per hydrogen atom, and using the fact that reionization is expected to happen during matter domination,
\begin{equation}
\tau \propto \int \ud z \, x_e \sqrt{1+z} \propto \int \ud[(1+z)^{3/2}] x_e.
\end{equation}
It is therefore handy to parameterize $x_e$ as a function of $y\equiv (1+z)^{3/2}$. As of March 2008 CAMB's default parameterization uses a $\tanh$-based fitting function\footnote{There was a sign typo in previous versions of this paper; thanks to Francesco Montanari for spotting it.}
\begin{equation}
x_e(y) = \frac{f}{2} \left[ 1 + \tanh\left(\frac{y(z_{re})-y}{\Delta_y}\right)\right],
\end{equation}
where $y(z_{re}) = (1+z_{re})^{3/2}$ is where $x_e = f/2$: i.e. $z_{re}$ measures where the reionization fraction is half of its maximum.  In other words, with this parameterization the optical depth agrees with that for an instantaneous reionization model at the same $z_{re}$ for all (matter-dominated) values of $\Delta_y$.
Except in early dark energy models this result is quite accurate for the expected range of $z_{re}$. In practice the input parameter is $\Delta_z$ and $\Delta_y$ is taken to be $1.5\sqrt{1+z_{re}}\Delta_z$.

\begin{figure}
\begin{center}
\epsfig{figure=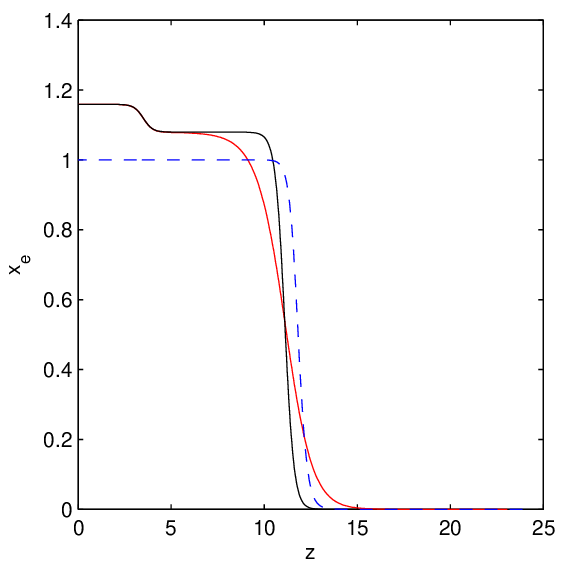,angle=0,width=8cm}
\epsfig{figure=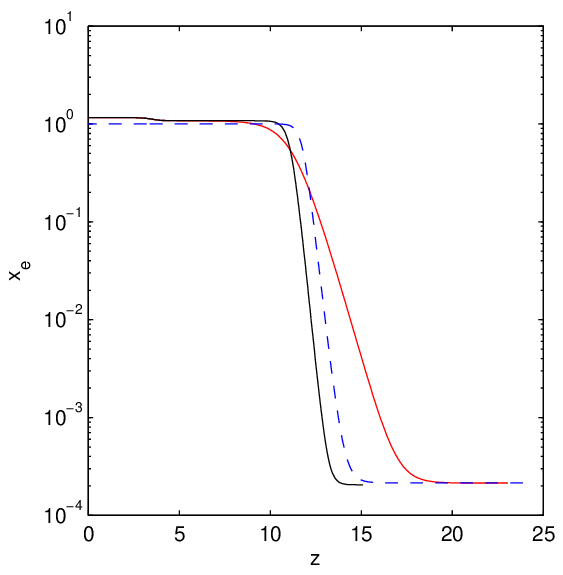,angle=0,width=8cm}
\caption{Three recombination histories all with $\tau = 0.09$. The dashed line is the model typically used by CMBFAST and CAMB prior to March 2008 with $f=1$.  The black line is the new model with $\Delta_z=0.5$, the red line with $\Delta_z=1.5$.
}
\label{reion}
\end{center}
\end{figure}

If hydrogen fully ionizes $f=1$. However the first ionization energy of helium is similar, and it is often assumed that helium first re-ionizes in roughly the same way. This is indeed seems to be the case in numerical simulations (see e.g. Ref.~\cite{Shapiro:2003gxa}). In this case $f = 1 + f_{He}$, where $f_{He}=n_{He}/n_H$ is easily calculated from the input helium mass fraction $Y_{He}$. This is CAMB's default value of $f$;  typically $f \sim 1.08$.

In addition at $z\sim 3.5$ helium probably gets doubly ionized. Due to the low redshift this only affects the optical depth by $\sim 0.001$, but for completeness this is included using a fixed tanh-like fitting function (modifying the above result for $\tau$ appropriately). Some reionization histories are shown in Fig.~\ref{reion}.

\bibliography{../../antony,../../cosmomc}

\end{document}